\documentclass[twocolumn,floats,showpacs,superscriptaddress,pre]{revtex4}
\usepackage{amsmath,amssymb,bm}
\usepackage{graphics}
\usepackage{epsfig}
\usepackage{graphicx}

\begin{document}

\title{Two-Time Quantum Mechanics }
\author{Natalia Gorobey, Alexander Lukyanenko}
\email{alex.lukyan@rambler.ru}
\affiliation{Department of Experimental Physics, St. Petersburg State Polytechnical
University, Polytekhnicheskaya 29, 195251, St. Petersburg, Russia}
\author{Inna Lukyanenko}
\email{inna.lukyanen@gmail.com}
\affiliation{Institut f\"{u}r Mathematik, TU Berlin, Strasse des 17 Juni 136, 10623
Berlin, Germany}

\begin{abstract}
A two-time quantum theory of a system of two particles with the
direct electromagnetic interaction based on a quantum version of the
action principle is considered. An analog of Schr\"{o}dinger
equation for the system is obtained.
\end{abstract}

\maketitle
\date{\today }

%\pacs{}

%\begin{multicols}{2}
%\narrowtext

%%%%%%%%%%%%%%%%%%%%%%%%%%%%%%%%%%%%%%%%%%%%%%%%%%%%%%%%%%%%%%%%%%%%%%%

%%\bigskip

\section{\textbf{INTRODUCTION}}

In the preprint \cite{GLL5} a generalized canonical form of a
multi-time dynamical theory was proposed. This form may be taken as
a basis for a new approach in quantum theory. This approach is based
on a so called quantum action principle (QAP) proposed in
\cite{GL1,GL2,GLL6}. In paper \cite{GLL5}, a two-time dynamical
system of two charges with a direct electromagnetic interaction was
considered
%%there
as a simple example. A quantum action operator for this system in
the first-order approximation of perturbation theory in a parameter
of the direct electromagnetic interaction was defined. In a
maximally simplified non-relativistic limit for this action, a
two-time integral, which corresponds to the Coulomb interaction
described, with relativistic accuracy, by the Feynman
retarded-advanced propagator \cite{S}, is conserved. It is this term
that makes impossible the ordinary canonical quantization procedure.

In the present work we develop a two-time quantum theory of the
relativistic Coulomb interaction of charges as a toy model for a
more complicated multi-time quantum electrodynamics. A two-time wave
equation, which is similar to
%%analogous to
Schr\"{o}dinger equation, is obtained.

\section{QUANTUM ACTION PRINCIPLE FOR TWO-TIME DYNAMICAL SYSTEM}

The action operator of a two-time dynamical theory has the form
\cite{GLL5}:
\begin{eqnarray}
\widehat{I} &=&\int\limits_{0}^{T_{1}}dt_{1}\overset{\cdot }{x}_{1k}\left(
t_{1}\right) \frac{\widetilde{\hslash }}{i}\frac{\delta }{\delta
x_{1k}\left( t_{1}\right) }  \notag \\
&&+\int\limits_{0}^{T_{2}}dt_{2}\overset{\cdot }{x}_{2k}\left( t_{2}\right)
\frac{\widetilde{\hslash }}{i}\frac{\delta }{\delta x_{2k}\left(
t_{2}\right) }  \notag \\
&&-\widehat{H}.  \label{1}
\end{eqnarray}%
Here $x_{1k}\left( t_{1}\right)$ and $x_{2k}\left( t_{2}\right) $
are trajectories of particles with own parameters of time, the dot
denotes the time derivative with respect to %%on
the time parameter corresponding to
each particle. The
constant $\widetilde{\hbar }$ is not equal to the ordinary Plank constant $%
\hbar $. The physical dimensionality of $\widetilde{\hbar }$ is $\left[ \widetilde{%
\hbar }\right] =Joule\cdot s^{2}$. A connection between two
constants will be introduced latter. A Hamiltonian part of the
action operator (\ref{1}) in the non-relativistic limit is:
\begin{eqnarray}
\widehat{H} &=&-\int\limits_{0}^{T_{1}}dt_{1}\frac{\widetilde{\hslash }^{2}}{%
2m_{1}}\frac{\delta ^{2}}{\delta x_{1}^{2}\left( t_{1}\right) }%
-\int\limits_{0}^{T_{2}}dt_{2}\frac{\widetilde{\hslash }^{2}}{2m_{2}}\frac{%
\delta ^{2}}{\delta x_{2}^{2}\left( t_{2}\right) }  \notag \\
&&+\frac{1}{2}e_{1}e_{2}\int\limits_{0}^{T_{1}}dt_{1}\int%
\limits_{0}^{T_{2}}dt_{2}\delta \left( s_{12}^{2}\right) ,  \label{2}
\end{eqnarray}%
where (velocity of light $c=1$)
\begin{equation}
s_{12}^{2}\equiv \left( t_{1}-t_{2}\right) ^{2}-\left( x_{1}-x_{2}\right)
^{2}  \label{3}
\end{equation}%
is the interval in the Minkowsky space. The action operator
(\ref{1}) is defined in a space of wave functionals $\Psi \left[
x_{1}\left( t_{1}\right) ,x_{2}\left( t_{2}\right) \right] $ where
in turn the variational derivatives
%%in one's turn
are defined as follows:
\begin{eqnarray}
\delta \Psi &\equiv &\int\limits_{0}^{T_{1}}dt_{1}\frac{\delta \Psi }{\delta
x_{1k}\left( t_{1}\right) }\delta x_{1k}\left( t_{1}\right)  \notag \\
&&+\int\limits_{0}^{T_{2}}dt_{2}\frac{\delta \Psi }{\delta x_{2k}\left(
t_{2}\right) }\delta x_{2k}\left( t_{2}\right)  \label{4}
\end{eqnarray}%
The Hamiltonian operator (\ref{2}) is formally Hermitian with respect to the
scalar product in the space of wave functionals:
\begin{eqnarray}
\left( \Psi _{1},\Psi _{2}\right) &\equiv &\int
\prod\limits_{t_{1}}d^{3}x_{1}\left( t_{1}\right)
\prod\limits_{t_{2}}d^{3}x_{2}\left( t_{2}\right)  \label{5} \\
&&\times \overline{\Psi }_{1}\left[ x_{1}\left( t_{1}\right) ,x_{2}\left(
t_{2}\right) \right] \Psi _{2}\left[ x_{1}\left( t_{1}\right) ,x_{2}\left(
t_{2}\right) \right] .  \notag
\end{eqnarray}%
The first two canonical terms of the action operator (\ref{1}) are
non-Hermitian, but we shall overcome this problem
%%the obstacle
by throwing away of corresponding imaginary parts of eigenvalues of
the action operator.

We formulate QAP as the eigenvalue problem for the action operator \cite%
{GLL6}:
\begin{equation}
\widehat{I}\Psi =\Lambda \Psi .  \label{6}
\end{equation}%
A solution $\Psi \left[ x_{1}\left( t_{1}\right) ,x_{2}\left( t_{2}\right) %
\right] $ of the equation (\ref{6}) describes a quantum state of motion of
the system. $\left\vert \Psi \left[ x_{1}\left( t_{1}\right) ,x_{2}\left(
t_{2}\right) \right] \right\vert ^{2}$ is a probability density of particles
movement along traejectories from a small neighbourhud of given trajectories
$x_{1k}\left( t_{1}\right) ,x_{2k}\left( t_{2}\right) $. It is the equation (%
\ref{6}) that is equivalent to Schr\"{o}dinger equation in the case
of a single-time dynamical theory \cite{GLL6}. In the case of
two-time dynamical theory considered here it will be reduced to a
two-time analog of the Schr\"{o}dinger equation.

It is useful to re-formulate the eigenvalue problem, introducing for
any wave functional $\Psi $ an auxiliary functional:
\begin{equation}
\Lambda \left[ x_{1}\left( t_{1}\right) ,x_{2}\left( t_{2}\right) \right]
\equiv \frac{\widehat{I}\Psi \left[ x_{1}\left( t_{1}\right) ,x_{2}\left(
t_{2}\right) \right] }{\Psi \left[ x_{1}\left( t_{1}\right) ,x_{2}\left(
t_{2}\right) \right] }.  \label{7}
\end{equation}%
Then QAP may be formulated in terms of a necessary condition that
the functional (\ref{7}) is equal to an eigenvalue of the action
operator. The following exponential representation
\begin{eqnarray}
&&\Psi \left[ x_{1}\left( t_{1}\right) ,x_{2}\left( t_{2}\right) \right]\equiv
\notag \\
&&\exp \left( \frac{i}{\widetilde{\hbar }}S\left[ x_{1}\left(
t_{1}\right) ,x_{2}\left( t_{2}\right) \right] +R\left[ x_{1}\left(
t_{1}\right) ,x_{2}\left( t_{2}\right) \right] \right)  \label{8}
\end{eqnarray}%
with real functionals $S\left[ x_{1}\left( t_{1}\right) ,x_{2}\left(
t_{2}\right) \right] ,R\left[ x_{1}\left( t_{1}\right) ,x_{2}\left(
t_{2}\right) \right] $ is useful for a wave functional, in
particular, for quasi-classical decomposition of a solution. At this
stage, analytic properties of the Coulomb potential are important.
In the work \cite{GL2}, QAP was formulated for the case of a
real-analytic potential. In order to use this concrete
formulation, let us regularize the Coulomb potential, replacing the $\delta $%
-function by the exponent as follows:
\begin{equation}
\delta \left( s_{12}^{2}\right) \rightarrow \frac{1}{\sqrt{2\pi \sigma }}%
\exp \left( -\frac{\left( s_{12}^{2}\right) ^{2}}{2\sigma }\right) .
\label{9}
\end{equation}%
At the final stage of calculations, the limit $\sigma \rightarrow 0$
is supposed. With this regularization the functionals $S\left[
x_{1}\left( t_{1}\right) ,x_{2}\left( t_{2}\right) \right] ,R\left[
x_{1}\left( t_{1}\right) ,x_{2}\left( t_{2}\right) \right] $ may be
looked for as real-analytic functionals of trajectories of
particles:
\begin{eqnarray}
S\left[ x_{1},x_{2}\right] &=&\int\limits_{0}^{T_{1}}dt_{1}\int%
\limits_{0}^{T_{2}}dt_{2}s\left( t_{1},t_{2},x_{1}\left( t_{1}\right)
,x_{2}\left( t_{2}\right) \right) ,  \label{10} \\
s &\equiv &s_{k;}^{\left( 10\right) }\left( t_{1},t_{2}\right) x_{1k}\left(
t_{1}\right) +s_{;k}^{\left( 01\right) }\left( t_{1},t_{2}\right)
x_{2k}\left( t_{2}\right)  \notag \\
&&+s_{kl;}^{\left( 20\right) }\left( t_{1},t_{2}\right) x_{1k}\left(
t_{1}\right) x_{1l}\left( t_{1}\right)  \notag \\
&&+s_{k;l}^{\left( 11\right) }\left( t_{1},t_{2}\right) x_{1k}\left(
t_{1}\right) x_{2l}\left( t_{2}\right)  \notag \\
&&+s_{;kl}^{\left( 02\right) }\left( t_{1},t_{2}\right) x_{2k}\left(
t_{2}\right) x_{2l}\left( t_{2}\right)  \notag \\
&&+...,  \label{11}
\end{eqnarray}
\begin{eqnarray}
R\left[ x_{1},x_{2}\right] &=&\int\limits_{0}^{T_{1}}dt_{1}\int%
\limits_{0}^{T_{2}}dt_{2}r\left( t_{1},t_{2},x_{1}\left( t_{1}\right)
,x_{2}\left( t_{2}\right) \right) ,  \label{12} \\
r &\equiv &r_{k;}^{\left( 10\right) }\left( t_{1},t_{2}\right) x_{1k}\left(
t_{1}\right) +r_{;k}^{\left( 01\right) }\left( t_{1},t_{2}\right)
x_{2k}\left( t_{2}\right)  \notag \\
&&+r_{kl;}^{\left( 20\right) }\left( t_{1},t_{2}\right) x_{1k}\left(
t_{1}\right) x_{1l}\left( t_{1}\right)  \notag \\
&&+r_{k;l}^{\left( 11\right) }\left( t_{1},t_{2}\right) x_{1k}\left(
t_{1}\right) x_{2l}\left( t_{2}\right)  \notag \\
&&+r_{;kl}^{\left( 02\right) }\left( t_{1},t_{2}\right) x_{2k}\left(
t_{2}\right) x_{2l}\left( t_{2}\right)  \notag \\
&&+....  \label{13}
\end{eqnarray}%
Substituting (\ref{10}) and  (\ref{12}) into (\ref{7}) and taking
into account (\ref{11}) and (\ref{13}), we obtain the auxiliary
functional $\Lambda \left[ x_{1}\left( t_{1}\right) ,x_{2}\left(
t_{2}\right) \right] $ as a functional series of particles
coordinates $x_{1k}\left( t_{1}\right) ,x_{2k}\left( t_{2}\right)$
degrees. Then the %%necessary
condition that the functional is equal to an eigenvalue of the
action operator leads to an infinite set of differential equations
for coefficients of the series (\ref{11}) and (\ref{13}). We do not
present here these equations (in the case of a single-time theory, see \cite%
{GLL6}). In the next section we shall obtain an equivalent formulation of
this necessary condition in terms of a two-time wave equation.

\section{TWO-TIME WAVE EQUATION}

In order to re-formulate QAP in terms of a wave equation, let us
introduce a multiplicative representation of a wave functional $\Psi
\left[ x_{1}\left( t_{1}\right) ,x_{2}\left( t_{2}\right) \right] $
in terms of a two-time wave function $\psi \left(
t_{1},t_{2},x_{1},x_{2}\right) $ as follows. Let us divide both time
intervals $\left[ 0,T_{1,2}\right] $ on $N_{1,2}$ small
intervals of an the same length $\varepsilon =T_{1,2}/N_{1,2}$ by points $%
t_{n_{1}}=n_{1}\varepsilon ,t_{n_{2}}=n_{2}\varepsilon $, and
approximate a trajectory of the system of two particles $\left(
x_{1k}\left( t_{1}\right) ,x_{2k}\left( t_{2}\right) \right) $ in a
configuration space by a broken line with vertices $\left(
x_{1k}\left( n_{1}\right) ,x_{2k}\left( n_{2}\right) \right)
,n_{1}=1,2,...,N_{1}$, $n_{2}=1,2,...N_{2}$, and the end
points $\left( x_{1k}\left( 0\right) ,x_{2k}\left( 0\right) \right) $ $%
\equiv \left( x_{1k}^{0},x_{1k}^{0}\right) $, $\left( x_{1k}\left(
N_{1}\right) ,x_{2k}\left( N_{2}\right) \right) $ $\equiv \left(
x_{1k}^{T_{1}},x_{1k}^{T_{2}}\right) $. Then the functionals
(\ref{10})and \ref{12}) may be approximated by corresponding
integral sums:
\begin{eqnarray}
S
&=&\sum\limits_{n_{1}=1}^{N_{1}}\sum\limits_{n_{2}=1}^{N_{2}}\varepsilon
^{2}s\left( t_{n_{1}}t_{n_{2}},x_{1}\left( n_{1}\right) ,x_{2}\left(
n_{2}\right) \right) ,  \label{14} \\
R
&=&\sum\limits_{n_{1}=1}^{N_{1}}\sum\limits_{n_{2}=1}^{N_{2}}\varepsilon
^{2}r\left( t_{n_{1}}t_{n_{2}},x_{1}\left( n_{1}\right) ,x_{2}\left(
n_{2}\right) \right) .  \label{15}
\end{eqnarray}%
The central point of passage from a wave functional to a wave
function is the following equality \cite{GL1}, \cite{GLL6}:
\begin{equation}
\widetilde{\hbar }=\varepsilon \hbar .  \label{16}
\end{equation}%
Taking into account (\ref{14}) and (\ref{15}), the exponential
representation of a wave functional (\ref{8}) can be transformed to
a product of values of the wave function
%%taken
at discrete moments of time in the power $\varepsilon $:
\begin{eqnarray}
\Psi \left[ x\right] &=&\prod\limits_{n_{1}=1}^{N_{1}}\prod%
\limits_{n_{2}=1}^{N_{2}}\psi ^{\varepsilon }\left(
t_{n_{1}}t_{n_{2}},x_{1}\left( n_{1}\right) ,x_{2}\left( n_{2}\right)
\right) ,  \label{17} \\
&&\psi ^{\varepsilon }\left( t_{n_{1}}t_{n_{2}},x_{1}\left( n_{1}\right)
,x_{2}\left( n_{2}\right) \right)\equiv   \notag \\
&&\exp \varepsilon \chi \left( t_{n_{1}}t_{n_{2}},x_{1}\left(
n_{1}\right) ,x_{2}\left( n_{2}\right) \right) ,  \label{18} \\
&&\chi \left( t_{n_{1}}t_{n_{2}},x_{1}\left( n_{1}\right) ,x_{2}\left(
n_{2}\right) \right)\equiv   \notag \\
&&
\frac{i}{\hbar }s\left( t_{n_{1}}t_{n_{2}},x_{1}\left( n_{1}\right)
,x_{2}\left( n_{2}\right) \right)  \notag \\
&&
+\varepsilon r\left( t_{n_{1}}t_{n_{2}},x_{1}\left( n_{1}\right)
,x_{2}\left( n_{2}\right) \right) .  \label{19}
\end{eqnarray}%
Here and further the product $\varepsilon r$ will be considered as a
single symbol. In this approximation the wave functional $\Psi $\ is a
function of coordinates of vertices $\left(
x_{1k}\left( n_{1}\right),x_{2k}\left( n_{2}\right) \right) $ of a
broken line. According to the
definition (\ref{4}), the variational derivatives of this wave functional
%%%???????????
must be replaced by the partial
derivatives as follows (\cite{GL1}):
\begin{eqnarray}
\frac{\delta \Psi }{\delta x_{1k}\left( t_{n_{1}}\right) } &\equiv &\frac{1}{%
\varepsilon }\frac{\partial \Psi }{\partial x_{1k}\left( n_{1}\right) }
\label{20} \\
&=&\frac{1}{\varepsilon }\sum\limits_{n_{2}=1}^{N_{2}}\varepsilon \frac{%
\partial \chi \left( t_{n_{1}}t_{n_{2}},x_{1}\left( n_{1}\right)
,x_{2}\left( n_{2}\right) \right) }{\partial x_{1k}\left( n_{1}\right) }\Psi
,  \notag
\end{eqnarray}
\begin{eqnarray}
\frac{\delta \Psi }{\delta x_{2k}\left( t_{n_{2}}\right) } &\equiv &\frac{1}{%
\varepsilon }\frac{\partial \Psi }{\partial x_{2k}\left( n_{2}\right) }
\label{21} \\
&=&\frac{1}{\varepsilon }\sum\limits_{n_{1}=1}^{N_{1}}\varepsilon \frac{%
\partial \chi \left( t_{n_{1}}t_{n_{2}},x_{1}\left( n_{1}\right)
,x_{2}\left( n_{2}\right) \right) }{\partial x_{2k}\left( n_{2}\right) }\Psi
.  \notag
\end{eqnarray}%
Then the action operator can be approximated by a differential
operator. The first term in the canonical part of the auxiliary
functional $\Lambda \left[ x_{1}\left( t_{1}\right) ,x_{2}\left(
t_{2}\right) \right] $ is approximated as follows:
\begin{eqnarray}
&&\int\limits_{0}^{T_{1}}dt_{1}\overset{\cdot }{x}_{1k}\left( t_{1}\right)
\frac{\widetilde{\hslash }}{i}\frac{\delta \ln \Psi }{\delta x_{1k}\left(
t_{1}\right) }  \label{22} \\
&\simeq &\sum\limits_{n_{1}=1}^{N_{1}}\varepsilon \frac{\hbar }{i}\frac{%
x_{1k}\left( n_{1}\right) -x_{1k}\left( n_{1}-1\right) }{\varepsilon }\frac{%
\partial \ln \Psi }{\partial x_{1k}\left( n_{1}\right) }  \notag \\
&\simeq &\frac{\hbar }{i}\sum\limits_{n_{1}=1}^{N_{1}}\varepsilon
\sum\limits_{n_{2}=1}^{N_{2}}\left[ \chi \left(
t_{n_{1}}t_{n_{2}},x_{1}\left( n_{1}\right) ,x_{2}\left(
n_{2}\right)
\right) \right.  \notag \\
&&\left. -\chi \left( t_{n_{1}-1},t_{n_{2}},x_{1}\left( n_{1}-1\right)
,x_{2}\left( n_{2}\right) \right) \right]  \notag \\
&&-\frac{\hbar }{i}\sum\limits_{n_{1}=1}^{N_{1}}\varepsilon
\sum\limits_{n_{2}=1}^{N_{2}}\left[ \chi \left(
t_{n_{1}}t_{n_{2}},x_{1}\left( n_{1}\right) ,x_{2}\left(
n_{2}\right)
\right) \right.  \notag \\
&&\left. -\chi \left( t_{n_{1}-1},t_{n_{2}},x_{1}\left( n_{1}\right)
,x_{2}\left( n_{2}\right) \right) \right]  \notag \\
&\simeq &\frac{\hbar }{i}\sum\limits_{n_{2}=1}^{N_{2}}\varepsilon
\left[ \chi \left( T_{1},t_{n_{2}},x_{1}^{T_{1}},x_{2}\left(
n_{2}\right) \right)
\right.  \notag \\
&&\left. -\chi \left( 0,t_{n_{2}},x_{1}^{0},x_{2}\left( n_{2}\right) \right)
\right]  \notag \\
&&-\frac{\hbar }{i}\sum\limits_{n_{1}=1}^{N_{1}}\varepsilon
\sum\limits_{n_{2}=1}^{N_{2}}\varepsilon \frac{\partial \chi \left(
t_{n_{1}}t_{n_{2}},x_{1}\left( n_{1}\right) ,x_{2}\left(
n_{2}\right) \right) }{\partial t_{1}\left( n_{1}\right) }.  \notag
\end{eqnarray}%
The second term in (\ref{7}) is approximated in the similar way. In the limit $%
\varepsilon \rightarrow 0$ the full canonical part of the auxiliary
functional $\Lambda \left[ x_{1}\left( t_{1}\right) ,x_{2}\left(
t_{2}\right) \right] $ can be written in an integral form:
\begin{eqnarray}
&&\frac{\hbar }{i}\int\limits_{0}^{T_{1}}dt_{1}\int\limits_{0}^{T_{2}}dt_{2}%
\left[ \frac{1}{T_{1}T_{2}}\left( \chi \left(
T_{1},t_{2},x_{1}^{T_{1}},x_{2}\left( t_{2}\right) \right) \right. \right.
\notag \\
&&\left. -\chi \left( 0,t_{2},x_{1}^{0},x_{2}\left( t_{2}\right) \right)
\right) +\frac{1}{T_{1}T_{2}}\left( \chi \left( t_{1},T_{2},x_{1}\left(
t_{1}\right) ,x_{2}^{T_{2}}\right) \right.  \notag \\
&&\left. -\chi \left( t_{1},0,x_{1}\left( t_{1}\right) ,x_{2}^{T_{2}}\right)
\right) -\frac{1}{T_{2}}\frac{\partial \chi \left( t_{1},t_{2},x_{1}\left(
t_{1}\right) ,x_{2}\left( t_{2}\right) \right) }{\partial t_{1}}  \notag \\
&&\left. -\frac{1}{T_{1}}\frac{\partial \chi \left( t_{1},t_{2},x_{1}\left(
t_{1}\right) ,x_{2}\left( t_{2}\right) \right) }{\partial t_{2}}\right] .
\label{23}
\end{eqnarray}
Let us consider the Hamiltonian part of the auxiliary functional $\Lambda %
\left[ x_{1}\left( t_{1}\right) ,x_{2}\left( t_{2}\right) \right] $. The
first term, which corresponds to the first free particle, is approximated as
follows:
\begin{eqnarray}
&&\frac{\widetilde{\hbar }^{2}}{2m_{1}}\int\limits_{0}^{T_{1}}dt_{1}\frac{%
\delta ^{2}\Psi }{\delta x_{1}^{2}\left( t_{1}\right) }\frac{1}{\Psi }
\label{24} \\
&\simeq &\frac{\hbar ^{2}}{2m_{1}}\sum\limits_{n_{1}=1}^{N_{1}}\varepsilon %
\left[ \left( \sum\limits_{n_{2}=1}^{N_{2}}\varepsilon
\frac{\partial \chi \left( t_{n_{1}}t_{n_{2}},x_{1}\left(
n_{1}\right) ,x_{2}\left( n_{2}\right)
\right) }{\partial x_{1k}}\right) ^{2}\right.  \notag \\
&&\left. +\sum\limits_{n_{2}=1}^{N_{2}}\varepsilon \frac{\partial
^{2}\chi \left( t_{n_{1}}t_{n_{2}},x_{1}\left( n_{1}\right)
,x_{2}\left( n_{2}\right) \right) }{\partial x_{1k}^{2}}\right]
\notag
\end{eqnarray}%
The second term, which corresponds to the second free particle, is
approximated in the similar way. In the limit $\varepsilon
\rightarrow 0$, the Hamiltonian part of the auxiliary functional
$\Lambda \left[ x_{1}\left( t_{1}\right) ,x_{2}\left( t_{2}\right)
\right] $ equals to (notice that the third term, which corresponds
to Coulomb interaction, remains unchanged in this limit):
\begin{eqnarray}
&&-\int\limits_{0}^{T_{1}}dt_{1}\int\limits_{0}^{T_{2}}dt_{2}\left[ \frac{%
\hbar ^{2}}{2m_{1}}\left( \frac{\partial \chi \left( t_{1},t_{2},x_{1}\left(
t_{1}\right) ,x_{2}\left( t_{2}\right) \right) }{\partial x_{1k}}\right.
\right.  \notag \\
&&\times \int\limits_{0}^{T_{2}}d\widetilde{t}_{2}\frac{\partial \chi \left(
t_{1},\widetilde{t}_{2},x_{1}\left( t_{1}\right) ,x_{2}\left( \widetilde{t}%
_{2}\right) \right) }{\partial x_{1k}}  \notag \\
&&\left. -\frac{\partial ^{2}\chi \left( t_{1},t_{2},x_{1}\left(
t_{1}\right) ,x_{2}\left( t_{2}\right) \right) }{\partial x_{1k}^{2}}\right)
\notag \\
&&+\frac{\hbar ^{2}}{2m_{2}}\left( \frac{\partial \chi \left(
t_{1},t_{2},x_{1}\left( t_{1}\right) ,x_{2}\left( t_{2}\right) \right) }{%
\partial x_{2k}}\right.  \notag \\
&&\times \int\limits_{0}^{T_{1}}d\widetilde{t}_{1}\frac{\partial \chi \left(
\widetilde{t}_{1},t_{2},x_{1}\left( \widetilde{t}_{1}\right) ,x_{2}\left(
t_{2}\right) \right) }{\partial x_{2k}}  \notag \\
&&\left. \left. -\frac{\partial ^{2}\chi \left( t_{1},t_{2},x_{1}\left(
t_{1}\right) ,x_{2}\left( t_{2}\right) \right) }{\partial x_{2k}^{2}}\right) %
\right]  \notag \\
&&+e_{1}e_{2}\int\limits_{0}^{T_{1}}dt_{1}\int\limits_{0}^{T_{2}}dt_{2}%
\delta \left( s_{12}^{2}\right) .  \label{25}
\end{eqnarray}%
Collecting together all parts of the auxiliary functional $\Lambda \left[
x_{1}\left( t_{1}\right) ,x_{2}\left( t_{2}\right) \right] $, one can write
it in the form:
\begin{equation}
\Lambda \left[ x_{1}\left( t_{1}\right) ,x_{2}\left( t_{2}\right) \right]
=\lambda +\int\limits_{0}^{T_{1}}dt_{1}\int\limits_{0}^{T_{2}}dt_{2}\widehat{%
W}\chi ,  \label{26}
\end{equation}%
where
\begin{eqnarray}
\lambda &=&\int\limits_{0}^{T_{1}}dt_{1}\int\limits_{0}^{T_{2}}dt_{2}f\left(
t_{1},t_{2}\right)  \label{27} \\
f\left( t_{1},t_{2}\right) &\equiv &\left[ \frac{1}{T_{1}}\left( s\left(
T_{1},t_{2},0,0\right) \right. \right. \left. -s\left( 0,t_{2},0,0\right)
\right)  \notag \\
&&+\frac{1}{T_{2}}\left( s\left( t_{1},T_{2},0,0\right) \right. \left.
-s\left( t_{1},0,0,0\right) \right)  \notag \\
&&+\frac{1}{2m_{1}}\left( s_{k}^{\left( 10\right) }\left( t_{1},t_{2}\right)
\int\limits_{0}^{T_{2}}d\widetilde{t}_{2}s_{k}^{\left( 10\right) }\left(
t_{1},\widetilde{t}_{2}\right) \right.  \notag \\
&&-\hbar ^{2}\varepsilon r_{k}^{\left( 10\right) }\left( t_{1},t_{2}\right)
\int\limits_{0}^{T_{2}}d\widetilde{t}_{2}\varepsilon r_{k}^{\left( 10\right)
}\left( t_{1},\widetilde{t}_{2}\right)  \notag \\
&&\left. -\hbar ^{2}\varepsilon r_{kl}^{\left( 20\right) }\left(
t_{1},t_{2}\right) \right)  \notag
\end{eqnarray}
\begin{eqnarray}
&&+\frac{1}{2m_{2}}\left( s_{k}^{\left( 01\right) }\left( t_{1},t_{2}\right)
\int\limits_{0}^{T_{1}}d\widetilde{t}_{1}s_{k}^{\left( 01\right) }\left(
\widetilde{t}_{1},t_{2}\right) \right.  \notag \\
&&-\hbar ^{2}\varepsilon r_{k}^{\left( 01\right) }\left( t_{1},t_{2}\right)
\int\limits_{0}^{T_{1}}d\widetilde{t}_{1}\varepsilon r_{k}^{\left( 01\right)
}\left( \widetilde{t}_{1},t_{2}\right)  \notag \\
&&\left. \left. -\hbar ^{2}\varepsilon r_{kl}^{\left( 02\right) }\left(
t_{1},t_{2}\right) \right) \right] +e_{1}e_{2}\delta \left( \left(
t_{1}-t_{2}\right) ^{2}\right)  \label{28}
\end{eqnarray}%
is an eigenvalue of the action operator, and
\begin{eqnarray*}
\widehat{W}\chi &\equiv &\frac{\hbar }{i}\left[ \frac{1}{T_{1}}\left( \chi
\left( T_{1},t_{2},x_{1}^{T_{1}},x_{2}\left( t_{2}\right) \right) \right.
\right. \\
&&\left. -\chi \left( 0,t_{2},x_{1}^{0},x_{2}\left( t_{2}\right) \right)
\right) \\
&&+\frac{1}{T_{2}}\left( \chi \left( t_{1},T_{2},x_{1}\left( t_{1}\right)
,x_{2}^{T_{2}}\right) \right. \\
&&\left. -\chi \left( t_{1},0,x_{1}\left( t_{1}\right) ,x_{2}^{T_{2}}\right)
\right) \\
&&-\frac{\partial \chi \left( t_{1},t_{2},x_{1}\left( t_{1}\right)
,x_{2}\left( t_{2}\right) \right) }{\partial t_{1}} \\
&&\left. -\frac{\partial \chi \left( t_{1},t_{2},x_{1}\left( t_{1}\right)
,x_{2}\left( t_{2}\right) \right) }{\partial t_{2}}\right] \\
&&-\left[ \frac{\hbar ^{2}}{2m_{1}}\left( \frac{\partial \chi \left(
t_{1},t_{2},x_{1}\left( t_{1}\right) ,x_{2}\left( t_{2}\right) \right) }{%
\partial x_{1k}}\right. \right. \\
&&\times \int\limits_{0}^{T_{2}}d\widetilde{t}_{2}\frac{\partial \chi \left(
t_{1},\widetilde{t}_{2},x_{1}\left( t_{1}\right) ,x_{2}\left( \widetilde{t}%
_{2}\right) \right) }{\partial x_{1k}} \\
&&\left. -\frac{\partial ^{2}\chi \left( t_{1},t_{2},x_{1}\left(
t_{1}\right) ,x_{2}\left( t_{2}\right) \right) }{\partial x_{1k}^{2}}\right)
\end{eqnarray*}
\begin{eqnarray}
&&+\frac{\hbar ^{2}}{2m_{2}}\left( \frac{\partial \chi \left(
t_{1},t_{2},x_{1}\left( t_{1}\right) ,x_{2}\left( t_{2}\right) \right) }{%
\partial x_{2k}}\right.  \notag \\
&&\times \int\limits_{0}^{T_{1}}d\widetilde{t}_{1}\frac{\partial \chi \left(
\widetilde{t}_{1},t_{2},x_{1}\left( \widetilde{t}_{1}\right) ,x_{2}\left(
t_{2}\right) \right) }{\partial x_{2k}}  \notag \\
&&\left. \left. -\frac{\partial ^{2}\chi \left( t_{1},t_{2},x_{1}\left(
t_{1}\right) ,x_{2}\left( t_{2}\right) \right) }{\partial x_{2k}^{2}}\right) %
\right]  \notag \\
&&+e_{1}e_{2}\int\limits_{0}^{T_{1}}dt_{1}\int\limits_{0}^{T_{2}}dt_{2}%
\delta \left( s_{12}^{2}\right) -f\left( t_{1},t_{2}\right)  \label{29}
\end{eqnarray}%
is a part of the auxiliary functional $\Lambda \left[ x_{1}\left(
t_{1}\right) ,x_{2}\left( t_{2}\right) \right] $ which depends only
on non-zero degrees of coordinates of particles. It is this part of
the auxiliary functional $\Lambda \left[ x_{1}\left( t_{1}\right)
,x_{2}\left( t_{2}\right) \right] $ that gives a two-time wave
equation, which corresponds QAP:
\begin{equation}
\widehat{W}\chi =0.  \label{30}
\end{equation}%
This equation must to be solved for an arbitrary trajectory $\left(
x_{1k}\left( t_{1}\right) ,x_{2k}\left( t_{2}\right) \right) $ of
the system with fixed end points.

Let us consider the limiting case, when the interaction of charges
is absent. In this case the wave equation (\ref{30}) has a simple
additive solution for the exponent $\chi $ of a wave function:
\begin{eqnarray}
\chi \left( t_{1},t_{2},x_{1},x_{2}\right)  &=&\chi _{1}\left(
t_{1},x_{1}\right) +\chi _{2}\left( t_{2},x_{2}\right) ,  \label{31} \\
\chi _{1}\left( t_{1},x_{1}\right)  &\equiv &-\frac{i}{\hbar }\frac{\left(
W_{1}t_{1}-p_{1k}x_{1k}\right) }{T_{2}},  \notag \\
\chi _{2}\left( t_{2},x_{2}\right)  &\equiv &-\frac{i}{\hbar }\frac{\left(
W_{2}t_{2}-p_{2k}x_{2k}\right) }{T_{1}},  \notag
\end{eqnarray}%
where the energy $W$ and the momentum $p_{k}$ of each particle obey the
ordinary relation:
\begin{equation}
W=\frac{p^{2}}{2m}.  \label{32}
\end{equation}%
In this case, the exponent (\ref{31}) defines a wave function of the
system as a product:
\begin{equation}
\psi \left( t_{1},t_{2},x_{1},x_{2}\right) =\psi _{1}\left(
t_{1},x_{1}\right) \psi _{2}\left( t_{2},x_{2}\right) .  \label{33}
\end{equation}
The energy of the two-time quantum dynamical system with the direct
interaction can be defined in the limit $T_{1}=T_{2}=T\rightarrow
\infty $ as follows:
\begin{equation}
W\equiv i\hbar \lim_{T\rightarrow \infty }\left( \frac{\partial \chi \left(
T,T,x_{1}^{T},x_{2}^{T}\right) }{\partial t_{1}}+\frac{\partial \chi \left(
T,T,x_{1}^{T},x_{2}^{T}\right) }{\partial t_{2}}\right) ,  \label{34}
\end{equation}%
at the condition that the limit does not depend on the end points $%
x_{1}^{T},x_{2}^{T}$. In the limiting case of free particles we have
as usually:
\begin{equation}
W=\frac{p_{1}^{2}}{2m_{1}}+\frac{p_{2}^{2}}{2m_{2}}.  \label{35}
\end{equation}

\section{\textbf{CONCLUSIONS }}

In conclusion, the new formulation of quantum mechanics based on a
quantum version of the action principle is equivalent to ordinary
Schr\"{o}dinger formulation of quantum mechanics in the case of a
single-time dynamical theory \cite{GLL6}. We have showed that the
proposed new approach leads to a two-time analog of the
Schr\"{o}dinger equation in the case of the two-time dynamical
theory of a two-particle system with the direct electromagnetic
interaction.

We thank V. A. Franke and A. V. Goltsev for useful discussions.

%%%\noindent $^{\ast }$ E-mail address: alex.lukyan@rambler.ru

%%%\noindent $^{+}$ E-mail address: inna.lukyan@mail.ru

%\begin{references}

\end{document}